\documentclass{ws-procs9x6}
\begin{document}

\title{The Use of the Scattering Phase Shift in Resonance Physics}

\author{M. Nowakowski and N. G. Kelkar}

\address{Departamento de Fisica, Universidad de los Andes, \\ 
A. A. 4976\\ 
Santafe de Bogota, Colombia \\
E-mail: mnowakos@uniandes.edu.co, nkelkar@uniandes.edu.co}

\maketitle

\abstracts{The scattering phase shift encodes a good 
amount of physical
information which can be used to study resonances from scattering data. 
Among others, it can be used to calculate the continuum density of states 
and the collision time in a resonant process. Whereas the first 
information can be employed to examine the evolution of unstable states 
directly from scattering data, the second one serves as a tool to detect 
resonances and their properties. We demonstrate both methods concentrating 
in the latter case on
`exotic' resonances in $\pi \pi$ and $\pi K$ scattering.}

\section{The Phase Shift and its Ambiguities}
The phase shift $\delta_l$ is a convenient variable to parameterize the 
scattering amplitude 
as known from many textbooks on quantum mechanics. In principle, the phase 
shift can 
be extracted from the differential cross section
$d\sigma/d\Omega$ 
or other observables. To understand a physical problem from a scattering
experiment it is often sufficient to know the 
differential cross section. A good example is a bump in the cross section which signals often
a resonance. However, the knowledge of the phase shift (or amplitude) can give us additional 
information on the dynamics of the scattering process, the properties of occurring resonances etc.
We shall discuss two such applications of the phase shift in the 
subsequent sections. Both have 
to do with resonances as intermediate states in the scattering process  
from which $\delta_l$ is determined. It is therefore not 
unimportant to remind the reader of some facts regarding the extraction of the phase shift
from experiment.
Only if the extraction is unambiguous and clean can we 
rely on the information stored in the 
extracted data. The usage of the differential cross section is not enough for this purpose because
of several ambiguities encountered in the process. Here we will just mention two of them: the 
so-called Yang \cite{yang} and Minami \cite{minami} ambiguities. The first one allows for two different 
solutions in case that only $S$ and $P$ waves are present (important) in the reaction under 
consideration. The relation between the two solutions, primed and unprimed, reads:
$\delta(S_{31})=\delta'(S_{31})$ and 
$\delta(P_{33})-\delta(P_{31})=\delta'(P_{31})-\delta'(P_{33})$.
The second famous ambiguity is called Minami after its discoverer. It is based as all such ambiguities
on the invariance of the differential cross section under a certain 
transformation. In case of the Minami 
ambiguity this is $\delta_{j+1/2} \leftrightarrow \delta_{j-1/2}$ where $j$ is the total angular
momentum. Hence more specifically one obtains
$\delta(S_1) \leftrightarrow  \delta(P_1)$, 
$\delta(P_3) \leftrightarrow  \delta(D_3)$ and
$\delta(D_5) \leftrightarrow \delta(F_5)$.
There are more ambiguities which we will not discuss here. But the point we wish to make is that
because of these ambiguities a differential cross section is not sufficient to extract the phase shift and 
more data such as the polarization observables
are necessary \cite{cence}.
In addition, the extracted phase shifts must 
pass a number of tests based on basic properties of the $S$-matrix like unitarity, analyticity 
and crossing symmetry. For instance, in $\pi\pi$ scattering these are gathered in the so-called Roy 
equations and the 
$\pi \pi$ phase shift has to satisfy the latter in order to be accepted as 
an unambiguous 
physical quantity \cite{kaminski}.

This way the experimental phase shifts are indeed trustworthy physical values which we can use to get 
some insight into the scattering process. In the next section we will give some examples on how these 
values might be used and processed.
 
\section{Physics from the Phase Shift}
Having convinced ourselves that the extraction of the phase shift is reliable, we proceed
to discuss two examples of how to use and interpret the derivative of the phase shift $\frac{d\delta_l}{dE}$
to obtain physically relevant information on the resonances occurring in the process.

\subsection{Continuum Level Density}
While calculating correction factors $B$ and $C$ to the equation of states 
of an ideal gas, namely, 
$pV=RT[1+B/V + C/V^2]$, Beth and Uhlenbeck \cite{beth} found that the derivative of the phase shift is
proportional to the difference of the density of states (of the outgoing particles) with and without 
interaction. In case of the  $l^{th}$ 
partial wave we have then, 
\begin{equation} \label {density}
n_l(E) -n^{(0)}_l(E) =\frac{2l+1}{\pi}
\frac{d\delta_l(E)}{dE}\, .
\end{equation}
This is an interesting result. To appreciate it, let us briefly recall the Fock-Krylov method
to study the time evolution of unstable states. It is based on the fact that unstable states cannot be 
eigenstates to the Hamiltonian and as a result we can expand the resonance states in terms of the
energy eigenstates; i.e.
\begin{equation} \label{Psi}
\vert \Psi \rangle =\int dE \,a(E) \vert E\rangle \,.
\end{equation}
With a little bit of algebra we can recast the survival amplitude in a 
Fourier transformation
of the so-called spectral function $\rho_{{ }_{\Psi}}=\vert a(A) \vert^2$ which is the probability density
to find the state $\vert E \rangle$ between $E$ and $E+dE$. This way we obtain the Fock-Krylov celebrated 
result \cite{fock}
\begin{equation} \label{fock}
A_{\Psi}(t)=\int_{E_{th.}}^{\infty}dE \rho_{{ }_{\Psi}}(E) e^{-iEt}
\end{equation}
used in many investigations on quantum time evolution \cite{fock2}.
Now the probability density and the continuum density of states are related by a constant. As long as
there are no interfering resonances $\frac{d\delta}{dE}$ is positive and the above identification
works without doubt. The best example is the Lorentzian (Breit-Wigner)
spectral density which is very often used in investigations of the decay problem \cite{raczka}. Indeed,
$T=\frac{\Gamma/2}{E_R -E -i\Gamma/2}$ gives rise to a phase shift $\delta$ such that
$\frac{d\delta}{dE}=\frac{\Gamma/2}{(E_R-E)^2 +\Gamma^2/4}$.
In case of several resonances the interference pattern can cause the derivative of the
phase shift to become negative. Therefore, it is safe to take the first resonance and neglect the subsequent
contribution of the higher lying resonances. We can do that if we want to study only the large time
behaviour of the time evolution because in this case due to the time-energy duality we need to know only 
the threshold behaviour of the phase shift.
We can parameterize  $\rho_{{ }_\Psi}$
without loss of generality in the form $\rho_{{ }_\Psi} \propto (E
-E_{\rm th.})^{\gamma}$ to account for the threshold. Hence we have
\begin{equation} \label{general1}
\rho_{{ }_{\Psi_l}}(E)
=G(E)(E-E_{\rm th.})^{\gamma (l)} \, .
\end{equation}
The form factor has the property $G(E) \to
0$ sufficiently fast as $E \to \infty$ and
a complex pole $z_{0}$, i.e. $1/G(z_{0})=0$ 
such that $\Im m(z_{0}) < 0$ and $\Re e(z_{0}) > 0$. This pole represents the
resonance at $E_R -i\Gamma/2$ and leads to the exponential decay. 
This generalizes the simple Breit-Wigner form which has the
pole but no threshold behaviour. 
These general properties of the spectral function allow 
us to compute
the survival probability by going to the complex plane.  
We choose the closed path
$C_{\rm R}= C_{\Im} + C_{\Re} + C_{\rm R}^{1/4}$, starting from
zero (after change of variables $y=E-E_{th}$) along the 
real axis ($C_{\Re}$) attaching to it a quarter of a circle with
radius ${\rm R}$ ($C_{\rm R}^{1/4}$) in the clockwise direction and completing 
the path by going upward the
imaginary axis up to zero ($C_{\Im}$). In the integral 
we let ${\rm R}$ go to infinity noting
that along $C_{{\rm R} \to \infty}^{1/4}$, the integral is zero. We subtract the contribution along the
imaginary axis. 
This gives
\begin{equation} \label{sum}
A_{\Psi_l}(t)=A_{\Psi_l}^{\rm E}(t) 
+ A_{\Psi_l}^{\rm P}(t)
\end{equation} 
with
\begin{equation} \label{expA}
A_{\Psi_l}^{\rm E}(t)=e^{-iE_{\rm th.}t}\lim_{{\rm R}\to \infty}
\oint_{C_{\rm R}}dz e^{-izt} z^{\gamma} G(z+ E_{\rm th.})=C_1\, 
e^{-iE_Rt}e^{-\Gamma/2\,t}
\end{equation}
by Cauchy's theorem and
\begin{eqnarray} \label{corrA}
A_{\Psi_l}^{\rm P}(t)&=&C_2 \,e^{-iE_{th.}t} 
\int_0^{\infty}dx e^{-xt} x^{\gamma}
G(-ix +E_{\rm th.}) \nonumber \\
&\simeq& C_2\, e^{-iE_{th.}t} G(E_{\rm th.})
\Gamma (\gamma +1)\, {1 \over t^{\gamma +1}}
\end{eqnarray}
\begin{figure}[h]
\epsfxsize=3cm
{\epsfxsize=1.9in\epsfbox{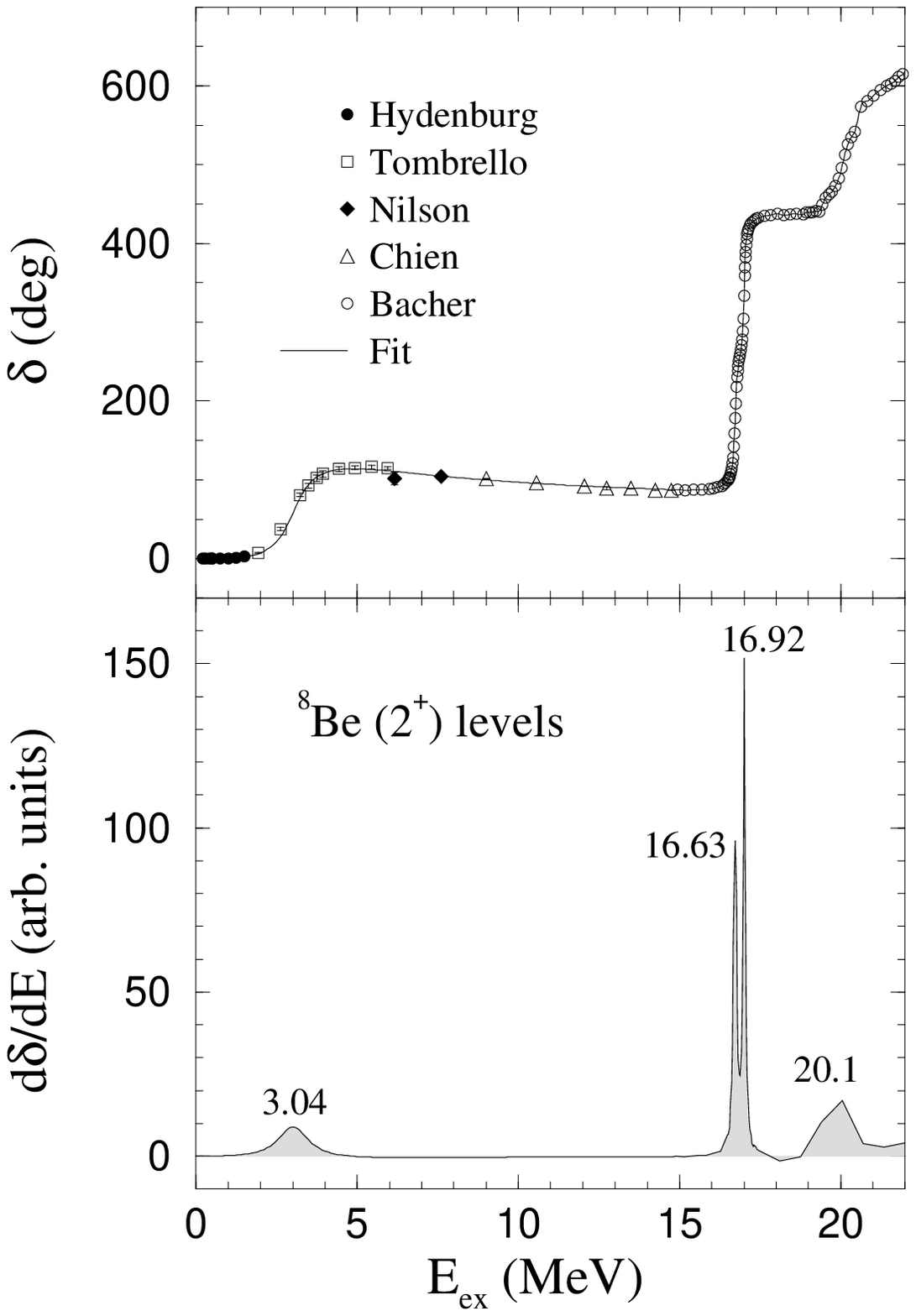}}\hspace{1cm}
{\epsfxsize=2.0in\epsfbox{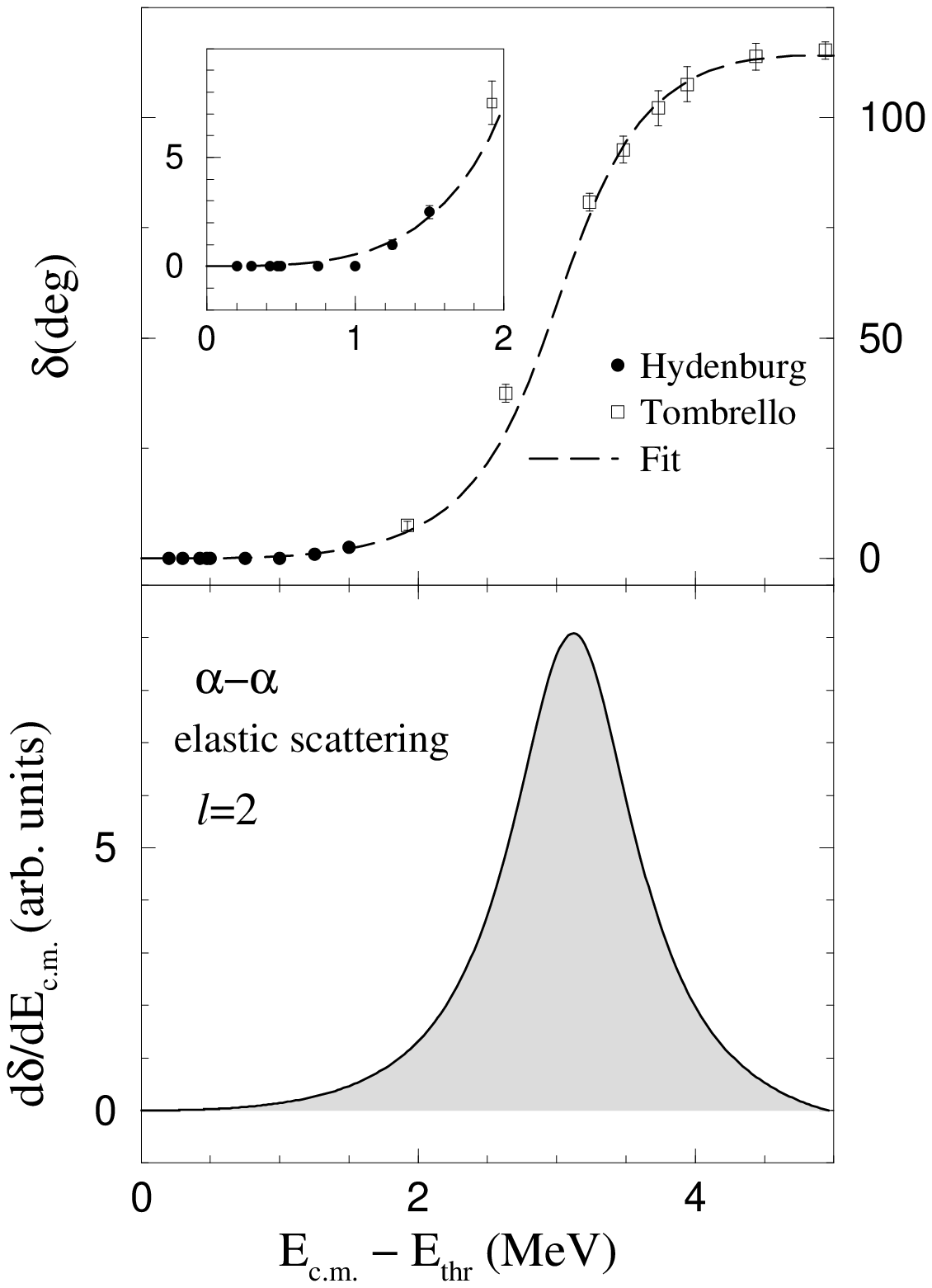}}
\caption{
D-wave phase shift (upper half) [9] and its derivative (lower half)
in 
$\alpha$-$\alpha$ elastic scattering as a function of
$E_{ex} = E_{c.m.}- E_{{ }^8{\rm Be}{\rm (ground state)}}$.
The figure on the right displays
the region of the first $2^+$ level of ${ }^8{\rm Be}$ in detail.
The inset displays the accuracy of our fit near the threshold energy
region which
is crucial for the large time behaviour of the decay law.} 
\end{figure}
for the integral along $C_{\Im}$, 
where the approximation is valid for large times $t$. 
In the above, $\Gamma (x)$ is the Euler's gamma function and
$C_1$ is a constant easily
calculable in terms of the parameters $E_R$, $\Gamma$ and $E_{\rm th.}$ and 
$C_2$ is $(-i)^{\gamma + 1}$. 
Equations (\ref{expA}) and
(\ref{corrA}) show the general features which we would expect: an
exponential decay law followed by inverse power law corrections. The latter is
independent of the details of $G(E)$ 
displaying also 
the dual nature of time and energy. 
Our choice of the elastic resonant reaction from which we want to
extract the information on the long tail of the decay is dictated by 
very good data at threshold. We opted for
$\alpha +\alpha \to { }^8Be(2^+) \to \alpha +\alpha$ (see Figure 1) \cite{data1}.
The analysis of this data following the mathematical method 
outlined above or 
alternatively by numerical integration reveals that the survival probability behaves as
$P_{\Psi}(t)=\vert A_{\Psi}(t)\vert^2 \sim t^{-6.36}$ for large times \cite{nonexp}.

Interestingly one can also use the interpretation of $d\delta/dE$ as continuum level density 
to calculate the density
of resonances per unit volume and unit invariant mass in a thermal environment e.g. in heavy
ion collisions \cite{broniowski}. This is  
$\frac{dn}{dM} \propto \int \frac{d^3p}{(2\pi)^3\pi}\, \frac{d\delta}{dM}\,
\frac{1}{\exp{(E_p/T)} \pm 1}$
and is proportional to the probability density to form this resonances.

\subsection{Time delay}
In Figure 1 we can see that all resonances  of ${ }^8Be$ with $l=2$ are nicely mapped
through the peaks of the derivative of the phase shift and the 
positions of these peaks 
correspond to masses of the resonant states. We would expect this due to the interpretation of
$d\delta/dE$ as a continuum level density. As explained above, 
such a level density should have a complex pole responsible for the 
exponential decay at the resonance position which, 
in the experimental data is reflected through a 
bump in the cross section.
If we want to map all resonances by this method, then it is more appealing to re-interpret
$d\delta/dE$ as a collision time or, in case this is positive, 
as a time delay in a
scattering experiment. Such an interpretation was pioneered by Wigner, Eisenbud and Bohm
\cite{wigner}
and is a topic of standard textbooks by now. For a wave-packet $A(E', E)$ centered around $E$
the exact expression is \cite{nuss}
\begin{equation} \label{time1}
\Delta t(E)= 8\pi^2 \int_0^{\infty}dE' \vert A(E',E)\vert^2 \,\, 2
\frac{d\delta}{dE'}
\end{equation}
which for a sufficiently narrow wave-packet $A(E',E)$ gives
\begin{equation} \label{time2}
\Delta t(E)=2\frac{d\delta}{dE}\, .
\end{equation}
With this interpretation we can reinforce the expectation that the
collision time $d\delta/dE$ peaks in the vicinity of a resonance (at the resonant energy to be exact).
Certainly, a collision is delayed if an intermediate state becomes on-shell (this happens
usually in the $s$-channel, but some curious examples of $t$-channel singularities also
exit \cite{NP}). 
We emphasize that the collision time (\ref{time2}) is strictly the 
difference between time spent with and without inetraction 
and not simply the time that a projectile spends in the scattering region of radius 
$a$. The latter in the presence of interaction 
reads \cite{nuss}
\begin{equation} \label{time3}
T(a)=\Delta t(E) + 8\pi^2 \int_0^{\infty} dE'\vert A(E',E) \vert^2
\left[2a -\frac{\sin2(\delta +k'a)}{k'}\right]
\end{equation}
and is positive definite in contrast to (\ref{time2}) which we expect to be 
positive at resonance energies. Equation (\ref{time3}) is only applicable if we can define 
unambiguously a radius $a$
which has a semi-classical character and/or assumes that the scattering particles have some
extension which, as the example of $e^+e^- \to Z^0 \to e^+e^-$ shows, is not always a good assumption
in resonant scattering. Furthermore, resonance production is per se due to interaction and therefore
(\ref{time2}) the right concept to use for our purposes.

We can now say that the survival amplitude 
(\ref{fock}) of a resonance is a Fourier 
transform of a collision time in momentum space if this resonance is an intermediate state in the process.

\section{Time delay and resonance physics}
Having identified the derivative of the phase shift as continuum level density and as time delay in 
resonant scattering, we can proceed to apply this concept to realistic examples 
(one of them is already displayed 
in Figure 1). It is, however, instructive to dwell first on some theoretical connections, misconceptions and
expectations. We note that we consider the usage of time delay in 
resonance physics as a supplementary tool to the other 
established methods. 

In literature one often encounters the statement of the correspondence 
`phase shift motion' 
$\leftrightarrow$ resonance.  Time delay is nothing else but the exact mathematical formulation of this
correspondence. However, this correspondence often carries a 
misunderstanding as it is attached also
to a $\pi$-jump of the phase shift. We stress that this $\pi$-jump is not a 
necessary condition for a resonance. In the spirit of time delay the 
condition is a peak around the resonance energy. 
Indeed, there are examples 
of prominent established resonances without the strong $\pi$-jump like 
$n + \alpha \to { }^5He(P_{1/2}) \to n + \alpha$ which is purely elastic 
with a jump from $0^o$ to $40^o$ `only' \cite{data2}.

A simple Breit-Wigner parameterization of the amplitude, i.e.
$T=\frac{\Gamma_i/2}{E_R-E -i\Gamma/2}$
corresponds to
$\delta =\frac{1}{2}\tan^{-1} \left[\frac{\Gamma_i(E_R-E)}
{(E_R-E)^2 +\Gamma^2/4 -\Gamma_i\Gamma/2}\right]$
which gives
$\left(\frac{d\delta}{dE}\right)_{E=E_R} 
=\frac{1}{\Gamma}\frac{Br}{Br-1/2}$.  This would mean that time delay is negative if
$Br < 1/2$.
An improvement can be reached by including a non-resonant 
background parametrized here by the diagonal phase $\zeta_i$ \cite{vacas}
and energy dependent width. One then gets
$\left(\frac{d\delta}{dE}\right)_{E=E_R} 
=\frac{1}{\Gamma(E_R)}\frac{Br(E_R)}{Br(E_R)-1/2}
+ \left(\frac{d\zeta_i}{dE}\right)_{E=E_R}$
which, in principle, can save the time delay from becoming negative
near a resonance.
However, we would not expect that 
when the resonant contribution is large.

Let us now confront this with experiment. In Figure 2 we have plotted the phase shift for the
$S_{11}$ resonances, the inelasticity parameter 
\begin{figure}[h]
\epsfxsize=3cm
\centerline{\epsfxsize=2.5in\epsfbox{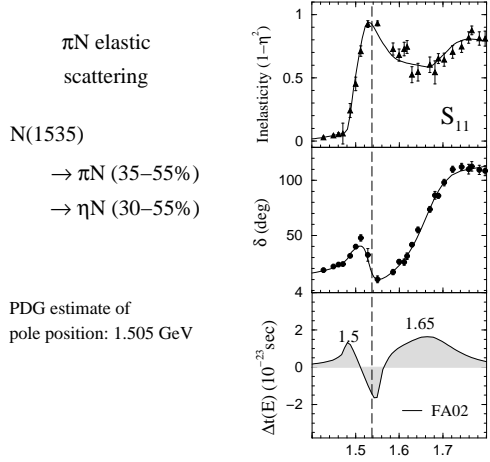}}
\caption{Single energy values of 
(a) inelasticities (triangles),
(b) phase shift (circles) and best fit curves to phase shift (solid lines)
in the $S_{11}$ partial wave of $\pi N$ elastic scattering 
and (c) distribution of time delay as a
function of energy available in the $\pi N$ centre of mass system.}
\end{figure}
(note that in case there are several channels, 
the $S$-matrix is written as $\eta \exp(2i\delta)$) and the 
time delay. 
First of all we find sharp peaks at $1.5$ GeV and $1.65$ GeV corresponding to the well known resonances
(Particle Data Group estimate of the pole value of the first $S_{11}$ resonance 
is $1.505$). Secondly, we get
these peaks in spite of 
the small branching ratio of $S_{11}(1535)$ which is $Br(\pi N)=35-55\%$ and $Br(\eta N) = 30-55\%$.
It is also clear that the time delay becomes negative when the inelasticity parameter is largest. This
can be understood as the loss of flux from the elastic channel due to the interpretation
of $\Delta t$ as density of states \cite{nee1}. In Figure 3 we have done a similar exercise for the
$P_{11}$ case \cite{baryon}. 
\begin{figure}[h]
\epsfxsize=3cm
\centerline{\epsfxsize=2.6in\epsfbox{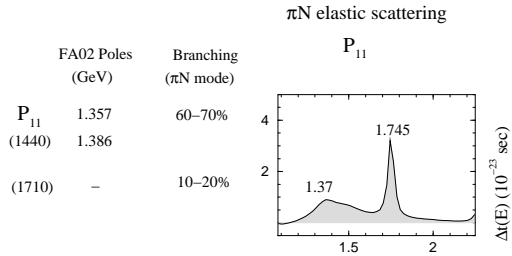}}
\caption{Time delay in the $P_{11}$ partial wave of $\pi N$ elastic 
scattering evaluated using the FA02 $T$-matrix solutions.}
\end{figure}
This is interesting from several points of view. Again we find two established resonances, 
but the focus is here on the three star $P_{11}(1710)$. We find this 
resonance by the time delay method at the right position
even if the $\pi$N branching ratio is as small as $10-20\%$. We find it 
by using the FA02 \cite{fa02} 
amplitudes even if the group which has performed the FA02 partial wave analysis cannot find the pole
corresponding to $P_{11}(1710)$. Whether or not this resonance exists is 
important for the theoretical prediction of the mass of the Pentaquark 
$\Theta (1540)$ \cite{pentaexp}.
Indeed, this prediction relies on the existence of the $P_{11}(1710)$ 
\cite{diakonov}. 
Through the time delay method
we find this resonance and also the Pentaquark \cite{penta1} at the right positions.

In passing we note that even resonances like $P_{13}(1585)$, $G_{17}(2190)$ and $H_{19}(2220)$ 
with $\pi$N branching ratios of $10-20\%$ leave 
clear fingerprints in the time delay plots \cite{baryon}.

\section{Resonances in $\pi \pi$ and $\pi K$ scattering}
The previous sections showed that the time delay method is reliable in 
nuclear and baryon resonance physics. We now turn our attention to the 
mesonic case \cite{meson}. 
To show how reliable the method indeed is
and how sensitive it is to small phase shift motion, we first apply the 
method to the case of the $\rho$-mesons. This
is depicted in Figure 4. 
\begin{figure}[h]
\epsfxsize=3cm
\centerline{\epsfxsize=1.9in\epsfbox{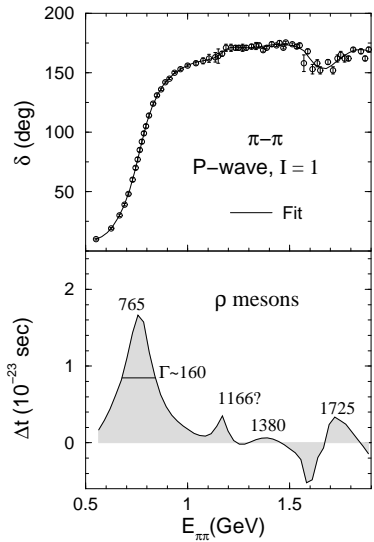}}
\caption{Time delay plot of the $\rho$ resonances evaluated from a
fit to the p-wave phase shifts in $\pi \pi$ elastic scattering.}
\end{figure}
Evidently, we find the `not-to-be-missed' $\rho(770)$, its first
excitation $\rho(1450)$ and its second excitation $\rho(1700)$ which are all indisputable resonances.
The peak at $1166$ MeV corresponds to a small phase motion and one could be tempted
to disregard it as a fluctuation. However, several other cases, 
among others the three star
resonance $D_{13}(1700)$ and the two star $F_{15}(2000)$, show that small 
phase shift `motion' can signify a resonance. This seems to be the case also here.
Particle Data Group lists also several mesons between $1100-2200$ MeV which 
by itself is not a remarkable fact. But at the recent Hadron 2001 
conference in Protvino 
some authors have pointed out a growing
evidence for a $\rho$-like resonance at $1200$ MeV which we think appears in our time delay plot\cite{protvino}. 
Such a state has been also predicted in a coupled channel quark/meson model\cite{dullemond}.
Our result in the $\rho$-meson sector is then an independent confirmation by the time delay method.

In the last few years the scalar sector attracted lots of attention. One of the reasons is the
`re-discovery' of the famous $\sigma$-meson and its `re-appearance' in the Particle Data Book.
The difficulty with this meson is reflected in the wide range of its possible mass, $400-1200$ MeV.
The time delay analysis for this sector is summarized in Figure 5. 
\begin{figure}[h]
\epsfxsize=3cm
\centerline{\epsfxsize=2.6in\epsfbox{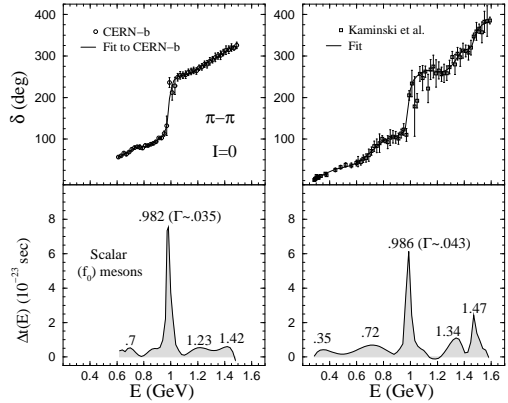}}
\caption{Time delay plot of the scalar meson resonances evaluated using
two different sets of the s-wave phase shifts [26,27]
in $\pi \pi$ elastic scattering.}
\end{figure}
Of course, $f_0(980)$ is a dominant
contribution here.
We identify the peak around $1.23 (1.34)$ GeV with $f_0(1370)$ for which Particle Data Group quotes
the range of possible pole mass between $1.2$ and $1.5$ GeV. Similarly the peak at $1.42 (1.47)$ GeV
is attributed to $f_0(1500)$ (the PDG value is $1.4-1.6$ GeV). The analysis of both phase shifts
reveals a resonance at $700$ MeV. If, in addition, we take the information of the Kaminski phase shift
we see also a peak at $350$ MeV. Can we take this as an evidence for two resonances? Let us first note that
in the region of $400-1200$ where the $\sigma$-meson is found, one can identify two accumulation points.
One at $350-600$ MeV and the other one at $700-850$ MeV. The low lying case is supported also by unitarized meson 
model\cite{rijken},
unitarized chiral perturbation theory \cite{oset1} and by unitarized quark model
\cite{uqm}, by coupled channel analysis\cite{locher} and by the so-called ABC effect \cite{abc}
which is with us since 1961 and the recent decay $J/\Psi \to \sigma \omega \to \pi\pi \omega$ \cite{jpsi}.
The $700$ MeV case finds its confirmation in Nambu-Jona-Lasinio models \cite{njl}, 
Weinberg's mended symmetry \cite{weinberg} and
Bethe-Salpeter calculation \cite{bethe}. 
Hence, these two accumulation points are not artificial constructs.      
They can be seen from the experimental results
the theoretical expectation\cite{more} so from the time delay  method.
Figure 6 serves as a cross check if the resonance found by the 
time delay method
in the elastic channel finds its confirmation also in other channels. 
As can be seen this is indeed the case. 

In the strange scalar sector the controversy regarding the
lightest scalar (called $\kappa$ meson) is even bigger. 
\begin{figure}[h]
\epsfxsize=3cm
\centerline{\epsfxsize=2.5in\epsfbox{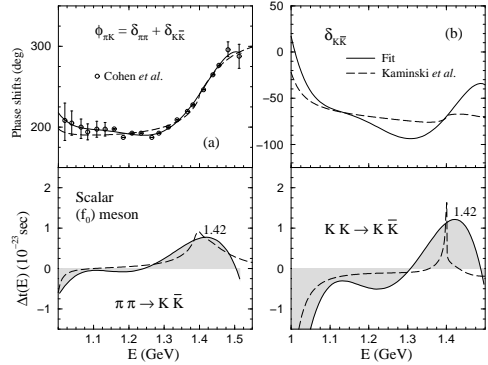}}
\caption{Energy dependence of the s-wave phase shifts and time delay
in the (a) $\pi \pi \rightarrow K \bar{K}$ and (b)
$K \bar{K} \rightarrow K \bar{K}$ reactions. The phase shifts
$\delta_{K \bar{K}}$ (solid lines in (b)) have been determined by
subtracting the fit to the
CERN-b $\pi \pi$ phase shift data from $\phi_{\pi K}$ in the
above figure (a). Dashed lines show the same quantities in the model
calculation of [40].}
\end{figure}
Figure 7 is an analysis of this
sector using two different phase shifts. This analysis also reveals the existence of two low lying 
resonances: one at $0.8$ GeV and the other around $1$ GeV which we identify 
with the putative $\kappa$-meson.

\begin{figure}[h]
\epsfxsize=3cm
\centerline{\epsfxsize=2.6in\epsfbox{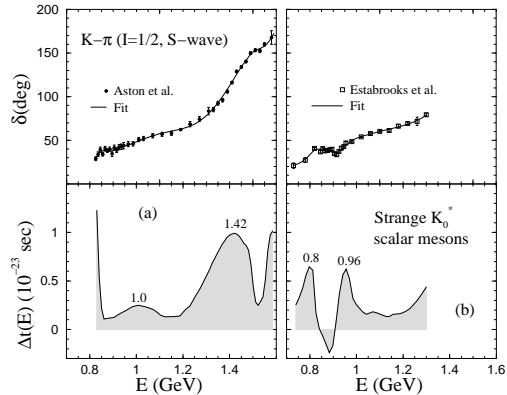}}
\caption{Time delay plots evaluated from the phase shifts in
s-wave, isospin 1/2 $K \pi$ scattering, displaying the strange scalar
mesons.}
\end{figure}
We have 
applied the time delay method
to many `standard' cases, the established baryon resonances, 
the $\rho$ mesons and the $K^*$- as well as
$K_2^*$-mesons (discussed in \cite{meson}) and found a good  agreement 
with data. Some less established resonances found by different methods get
confirmed through the time delay method. 
By using the $KN$ phase shift we found the recently discovered 
Pentaquark with a 
mass very close to the observed and predicted value \cite{penta1}. 
We found the spin-orbit partners of this Pentaquark
very close to the theoretical expectations \cite{penta2}, \cite{penta3}. 
Last but not least, our nuclear physics case 
discussed here in section two, shows also the virtues of the time delay 
method not only in finding nuclear levels, but also in studying the quantum evolution
of unstable systems for large times.


\begin{thebibliography}{0}
\bibitem{data1}
For an exhaustive list of phase shift data in the elastic $\alpha$-$\alpha$ scattering via
${ }^8 Be$ resonance see reference \cite{nonexp}.
\end{thebibliography}
\end{document}